\documentclass[aps,prl,twocolumn,amsmath,amssymb,showpacs,prb,superscriptaddress]{revtex4-1} 
\usepackage{epsf}   
\usepackage{graphicx}
\usepackage{gensymb}
\usepackage{sidecap}

\makeatletter

\@ifundefined{textcolor}{}
{%
 \definecolor{BLACK}{gray}{0}
 \definecolor{WHITE}{gray}{1}
 \definecolor{RED}{rgb}{1,0,0}
 \definecolor{GREEN}{rgb}{0,1,0}
 \definecolor{BLUE}{rgb}{0,0,1}
 \definecolor{CYAN}{cmyk}{1,0,0,0}
 \definecolor{MAGENTA}{cmyk}{0,1,0,0}
 \definecolor{YELLOW}{cmyk}{0,0,1,0}
}

\usepackage{color}\usepackage{url}\usepackage{epsf}

\begin{document}

\title {Unusual temperature dependence of band dispersion in Ba(Fe$_{1-x}$Ru$_x$)$_2$As$_2$ and its consequences for antiferromagnetic ordering}

\author{R.~S.~Dhaka}
\affiliation{Ames Laboratory and Department of Physics and Astronomy, Iowa State University, Ames, Iowa 50011, USA}
\author{S.~E.~Hahn}
\affiliation{Ames Laboratory and Department of Physics and Astronomy, Iowa State University, Ames, Iowa 50011, USA}
\author{E.~Razzoli}
\affiliation{Swiss Light Source, Paul Scherrer Institut, CH-5232 Villigen PSI, Switzerland}
\author{Rui~Jiang}
\affiliation{Ames Laboratory and Department of Physics and Astronomy, Iowa State University, Ames, Iowa 50011, USA}
\author{M.~Shi}
\affiliation{Swiss Light Source, Paul Scherrer Institut, CH-5232 Villigen PSI, Switzerland}
\author{B.~N.~Harmon}
\affiliation{Ames Laboratory and Department of Physics and Astronomy, Iowa State University, Ames, Iowa 50011, USA}
\author{A.~Thaler}
\author{S.~L.~Bud'ko}
\author{P.~C.~Canfield}
\author{Adam~Kaminski}
\email {kaminski@ameslab.gov}
\affiliation{Ames Laboratory and Department of Physics and Astronomy, Iowa State University, Ames, Iowa 50011, USA}

\date{\today}                                

\begin{abstract}
 We  have performed detailed studies of the temperature evolution of the electronic structure in Ba(Fe$_{1-x}$Ru$_x$)As$_2$ using Angle Resolved Photoemission Spectroscopy (ARPES). Surprisingly, we find that the binding energy of both hole and electron bands changes significantly with temperature in both pure and Ru substituted samples. The hole and electron pockets are well nested at low temperature in unsubstituted (BaFe$_{2}$As$_{2}$) samples, which likely drives the spin density wave (SDW) and resulting antiferromagnetic order. Upon warming, this nesting is degraded as the hole pocket shrinks and the electron pocket expands. Our results demonstrate that the temperature dependent nesting may play an important role in driving the antiferromagnetic/paramagnetic phase transition.

\end{abstract}

\maketitle

The phase transition from antiferromagnetic to paramagnetic order is believed to play a crucial role in the emergence of high temperature superconductivity in iron arsenic high temperature superconductors\cite{Kamihara08,ChenNature08,Chen08,ChenPRL08,Takahashi08}. The associated transition temperature can be controlled by tuning parameters such as chemical substitution  \cite{Rotter08,SefatPRL08,SasmalPRL08,CanfieldRev10,Johnston10,Thaler10,Hodovanets11,JiangKlintberg,Zhang09,SharmaSchnelle,Tropeano10,Rullier10} or pressure \cite{Paulpressure,Kimber09,Colombier09}. The superconducting dome exists in the phase diagram surrounding the point, where the AFM/PM phase transition temperature intersects T=0 axis and it is believed that the AF fluctuations are the glue pairing the electrons. Understanding the microscopic origin of this phase transition and the role of electronic structure is a prerequisite to validating the pairing mechanism and will likely aid in search for new superconducting materials. Our earlier studies \cite{Chiang2D} indicated that the AFM/PM phase transition is associated with a  two- to three-dimensional change of the electronic structure. Fermi surface nesting is believed to play an important role in the emergence of the SDW and AF order, however detailed understanding is still lacking. Data from chemical substitution yields only a partially consistent picture here. It is now well established that substitutions of Fe with Co, Ru, Ni or Pd increases the carrier concentration and shift the chemical potential higher (see for example Refs. \onlinecite{CanfieldRev10, LiuPRBNP}). This of course would worsen the FS nesting conditions, and naturally lead to weakening of the SDW. The case of isoelectronic substitution is much less clear. It is still debated whether or not substitution of As with isoelectronic P changes the carrier concentration \cite{FinkP, FengP}, however there is sufficient evidence for changes in dispersion of individual bands, which may affect nesting in a way similar to the previous case. A curious case is substitution of Fe with isoelectronic Ru, which results in a phase diagram similar to other substitutions (albeit for much higher Ru concentrations), but it was demonstrated to have negligible effects on the band dispersion up to ~40$\% $\cite{DhakaPRL11}. It appears that the SDW in this case is suppressed without significant change of the nesting condition and we speculated that magnetic dilution by less magnetic Ru may play a role in such case \cite{DhakaPRL11}.  

In ordinary materials, in absence of phase transitions, temperature has negligible effect on the band structure, and is mostly limited to changing the occupation close to the chemical potential (Fermi function) and changes of the quasiparticle lifetime reflected in the width of the spectral function measured by ARPES.  

\begin{figure*}
\includegraphics[width=7.2in]{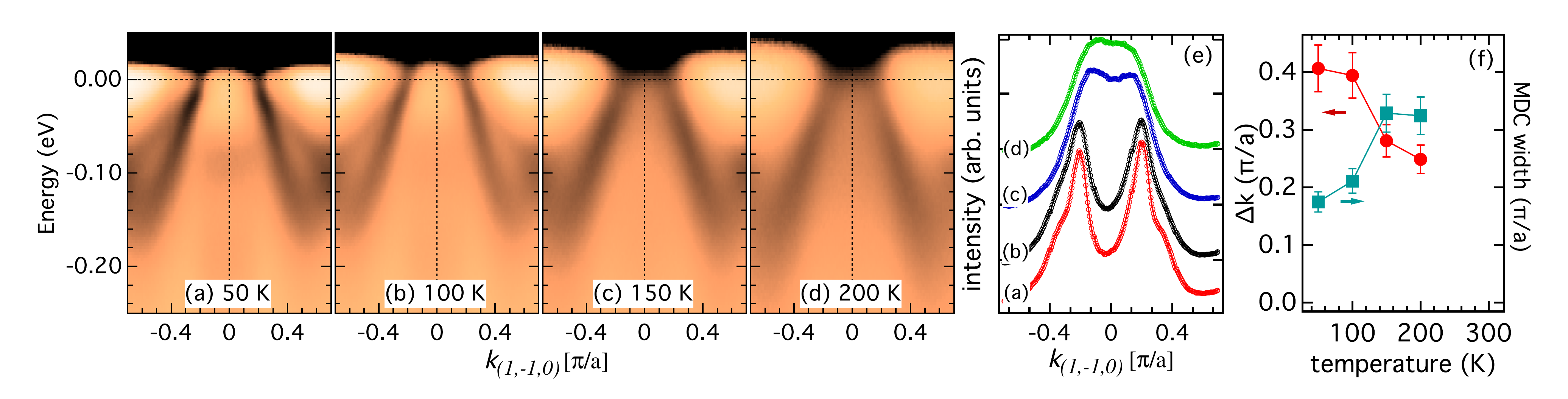}
\includegraphics[width=7.2in]{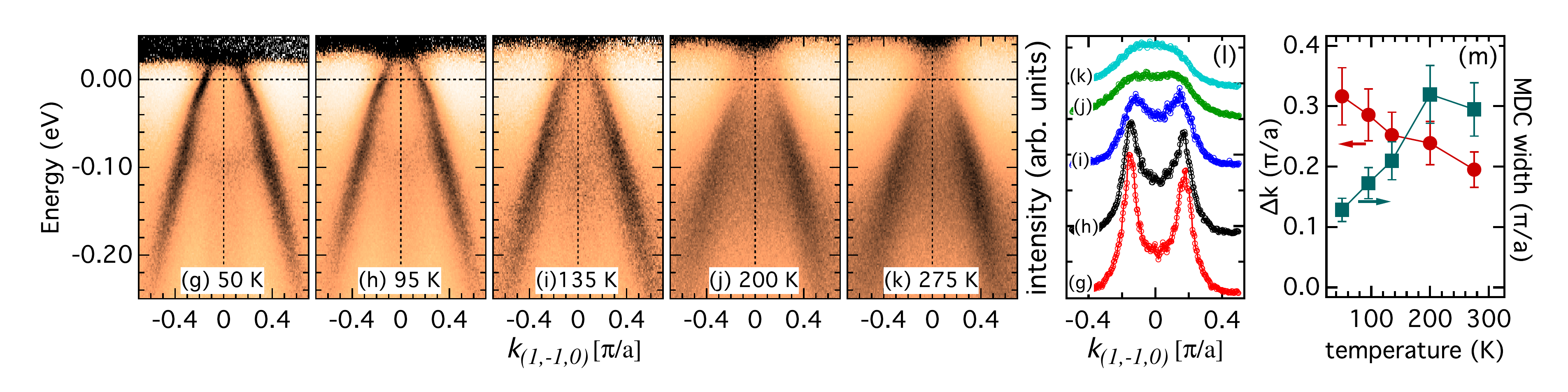}
\caption{(Color online) The band dispersion data of the parent BaFe$_2$As$_2$ compound measured with (a--f) h$\nu=35$~eV (Linear horizontal polarization) and (g--m) h$\nu=63$~eV (Linear vertical polarization) for various sample temperature. (e,l) The momentum distribution curves (MDCs) at $E_{\rm F}$ and (f,m) the extracted $\Delta k_{\rm F}$ values compared with sample temperature. }
\label{fig1}
\end{figure*}

In this Letter, we demonstrate that the band structure of one of the prototypical iron arsenic materials is changing significantly with temperature; in both pure and Ru substituted BaFe$_2$As$_2$ samples we observe significant shifts in the energy position of the bands. Upon increasing temperature the hole band moves to lower binding energy and eventually sinks below E$_f$ close to room temperature. The electron band, on the other hand, moves to lower binding energies with increasing temperature. This rather strange behavior destroys the nesting present at low temperatures upon warming the sample. Increasing temperature has therefore very similar effect to non-isoelectronic chemical substitution, where the nesting is lifted by changes of the chemical potential. Our finding explains the nearly linear decrease of the AFM transition temperature with increasing chemical substitution. It also points to a scenario where the AFM transition is driven by weakening of nesting with temperature rather than thermal fluctuations alone. 

Single crystals of Ba(Fe$_{1-x}$Ru$_x$)$_2$As$_2$ were grown out of self-flux using conventional high-temperature solution growth techniques\cite{Thaler10}, where FeAs and RuAs were synthesized in the same manner as in Ref~\onlinecite{Ni08}. The ARPES measurements were performed at 10.0.1 beamline (linear horizontal polarization, LHP) of the Advanced Light Source (ALS), Berkeley, California and at the SIS beamline (linear vertical polarization, LVP) of Swiss Light Source (SLS), Switzerland, using a Scienta R4000 electron analyzer in ultrahigh vacuum below $4\times10^{-11}$ mbar and at various sample temperatures. The data were also acquired using a laboratory based system consisting of a Scienta SES2002 electron analyzer and a GammaData helium ultraviolet lamp where we use the He~II line (h$\nu=40.8$~eV) for the band dispersion maps.  The samples were mounted on an Al pin using UHV compatible epoxy and {\it in situ} cleaved  perpendicular to the $c$-axis, yielding single layer surfaces in the $a$-$b$ plane. The measurements carried out on several samples and temperature cycling yielded similar results for the band dispersion and Fermi surface. The energy and angular resolution was set at 15~meV and ~$0.3^{\circ}$, respectively. High symmetry points were defined the same way as in Ref.~\onlinecite{LiuPRBNP}. {\it Ab initio} calculations were performed with VASP  \cite{Kresse93,Kresse96} using PAW pseudo-potentials  \cite{Blochl,Kresse99} and utilizing the LDA and PBE (Perdew-Burke-Ernzerhof) exchange-correlation functionals\cite{Perdew}. Experimental lattice parameters of $a=3.9624$, $c=13.0168$ and  $z_{\rm As}=0.3545$ are used from Ref.~\onlinecite{RotterPRB08}.  

\begin{figure}
\includegraphics[width=3.5in]{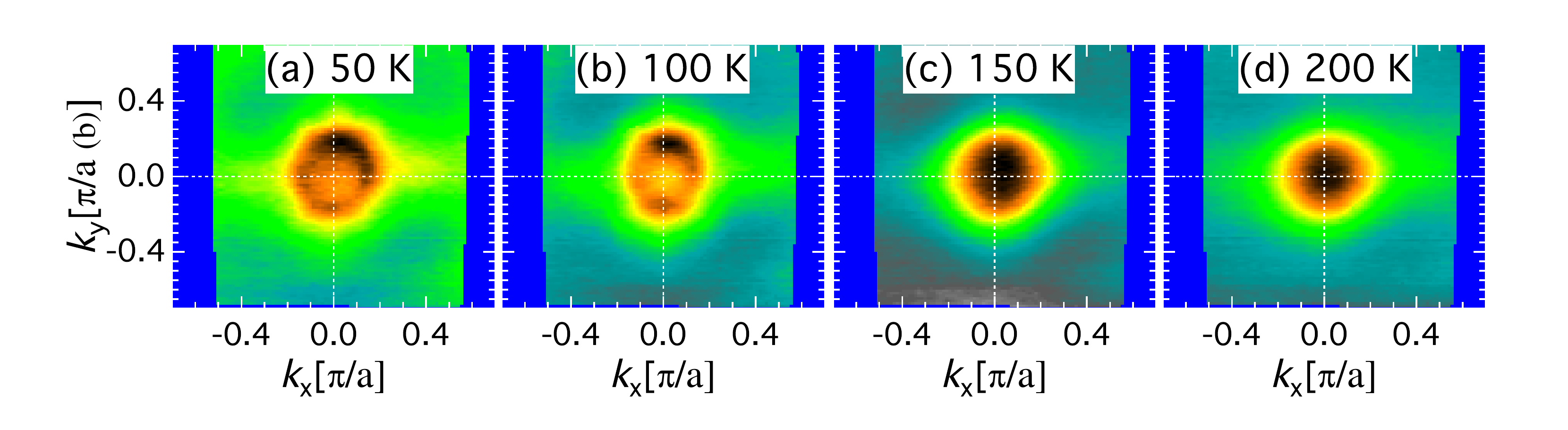}
\caption{(Color online) (a--d) The FS maps of BaFe$_2$As$_2$ measured with h$\nu=35$~eV (LHP) at various sample temperatures.}
\label{fig2}
\end{figure}

Figs. ~1(a--d) show the band dispersion maps for the parent compound BaFe$_2$As$_2$ measured at various temperatures using 35~eV ($k_{\rm z}\simeq2\pi/c$) photon energy {\it i.e.} the upper edge of the first Brillouin zone (${\rm Z}$ point)\cite{KondoPRB10}.  The second set of measurements were performed  using 63~eV photon energy and the results are shown in Figs.~1(g-m). All data were divided by resolution broadened Fermi functions to reveal band behavior in the proximity to the chemical potential. At low temperature the hole band at the center of the BZ crosses the chemical potential at roughly $\pm$ 0.2 $\pi/a$ resulting in hole like Fermi surface sheet (see also Fig. 2). With increasing temperature this band moves down, to larger binding energies and around 200~K the top of the band moves very close to the chemical potential.  Corresponding momentum distribution curves (MDC's), radii of the FS sheets and MDC widths are plotted in panels on the right. Here the two well separated MDC peaks present at low temperature move closer together and merge into flat top peak signifying significant (factor of ~2) reduction of the size of the hole pocket. Those changes are accompanied by increase in the MDC width, evident from raw data as well results of fitting in panels f and m. Whereas the increase of the MDC width may be associated with paramagnetic/antiferromagnetic-orthorhombic(tetragonal/orthorhombic) transition, the downward movement of the band seems to occur even above that temperature. In order to visualize the changes of the hole FS sheet we plot the ARPES intensity maps measured at the chemical potential in Fig. 2. The circular shape of this sheet present at low temperature shrinks and becomes a spot as the top of the band approaches the value of the chemical potential. It is clear from these data that size of this FS sheet decreases with increasing temperature, which will affect nesting in a very profound way. The above changes are strongest close to the top and bottom edges of the BZ along the k$_z$ direction. In Fig. 3 we plot the ARPES intensity along the (1,1,0) direction as a function k$_z$ momentum. The large hole pocket present at bottom and top edge of the zone (k$_z\sim18~\pi/c$ ) shrinks with temperature, whereas the dispersion close to center of the zone remains mostly unaffected. 

\begin{figure}
\includegraphics[width=3.2in]{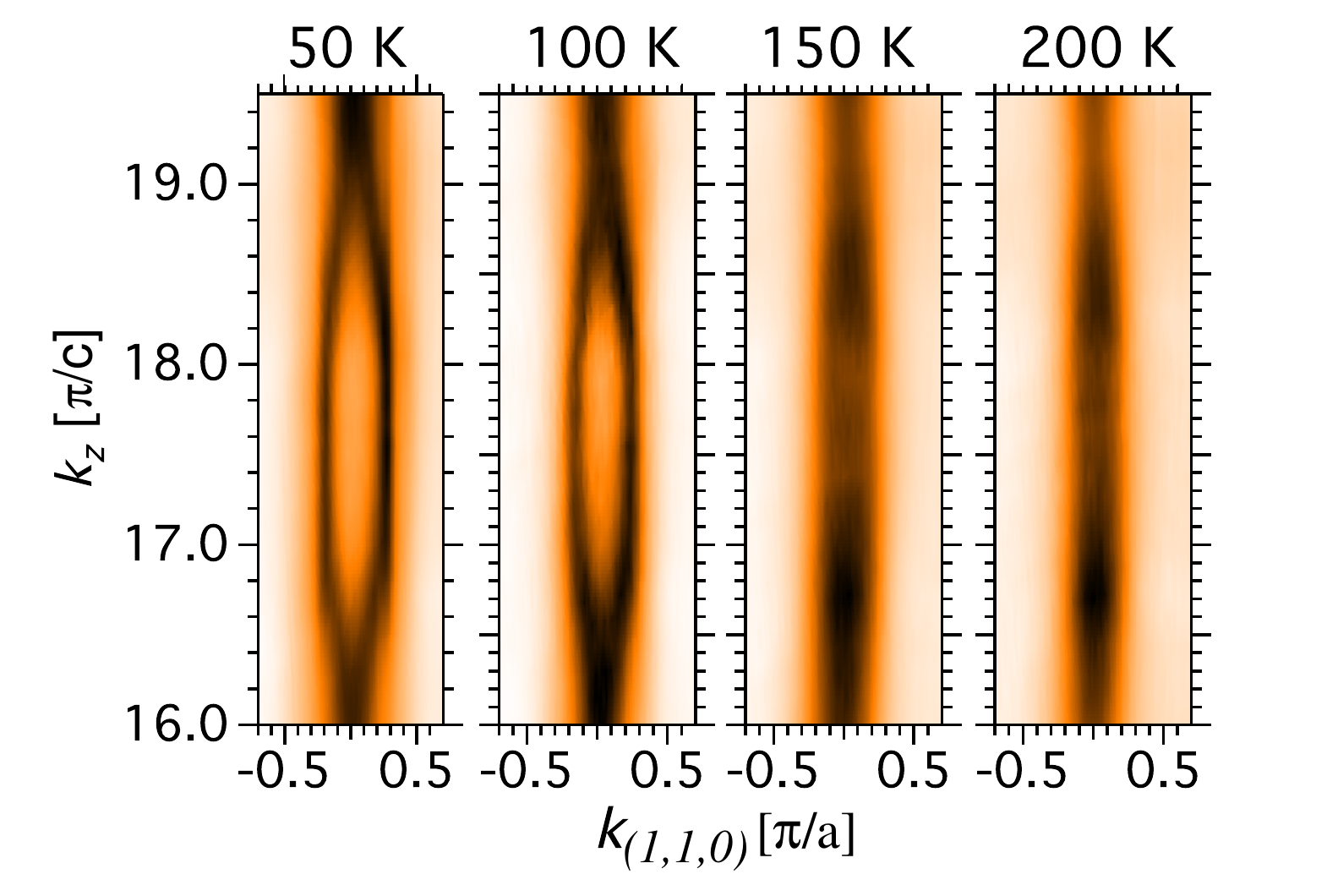}
\caption{(Color online) The hole Fermi surface maps of BaFe$_2$As$_2$ (at various sample temperatures) along the $k_{||}-k_z$  plane measured around the center of the BZ by changing h$\nu$.}
\label{fig3}
\end{figure}

\begin{figure}
\includegraphics[width=3.5in]{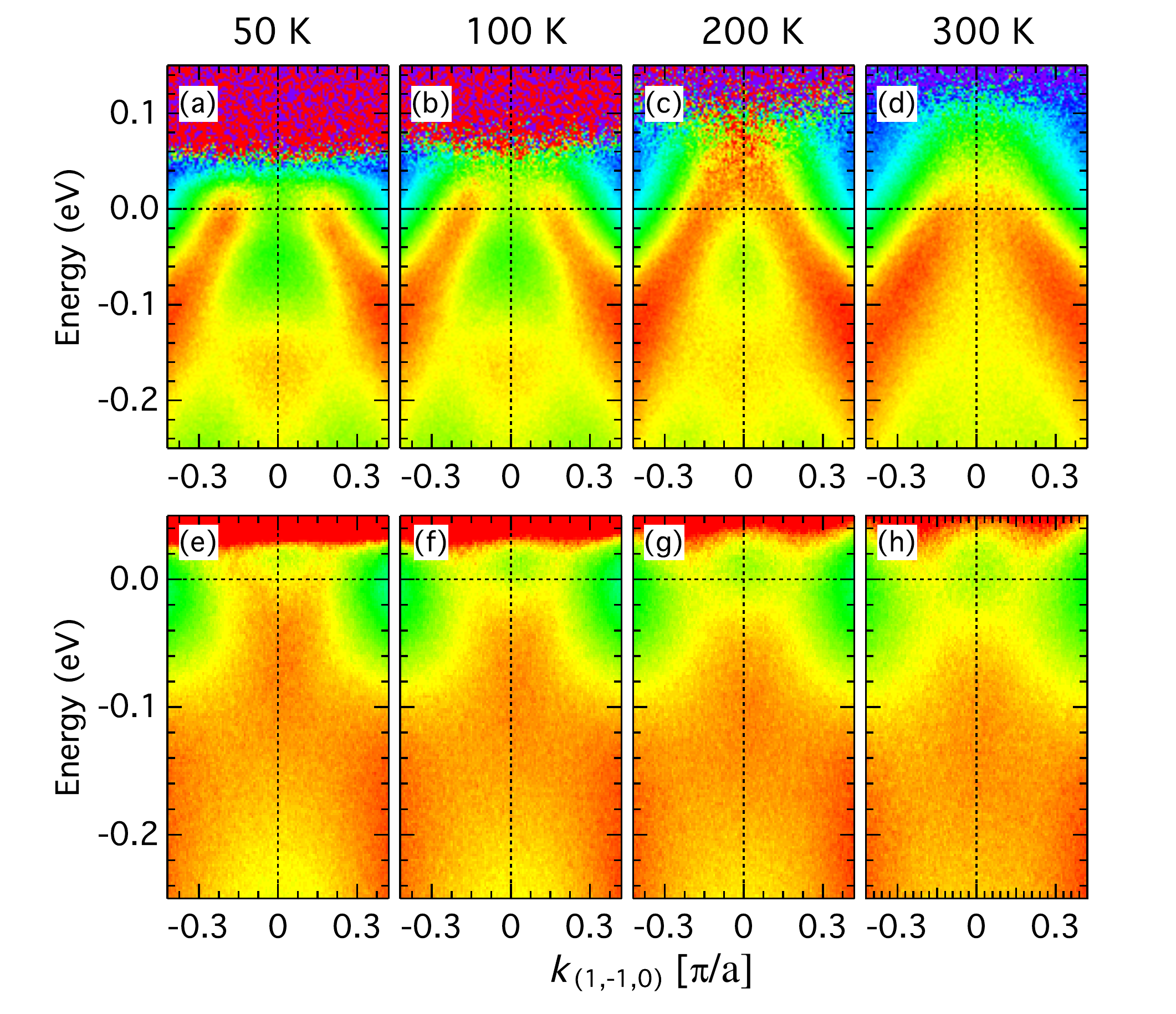}
\caption{(Color online) The band dispersion plots for Ba(Fe$_{1-x}$Ru$_x$)$_2$As$_2$, $x=0.36$ sample as a function of temperature. (a--d) hole pocket at the center of the BZ.a (e--h) electron pocket at the planar corner of the BZ. Data measured with photon energy h$\nu=40.8$~eV.}
\label{fig4}
\end{figure}

In the pure BaFe$_2$As$_2$ samples the band structure at low temperatures will of course be affected by the presence of the antiferromagnetic order. Since the electronic structure of Ru substituted samples, for x$\le$0.40 are almost identical\cite{DhakaPRL11}, we can access low temperatures in the paramagnetic state by choosing finite x values with suppressed magnetic/structural phase transitions\cite{Thaler10}. To verify that the temperature induced changes of the band structure are not caused by an AFM phase transition, we performed similar measurements using Ba(Fe$_{1-x}$Ru$_x$)$_2$As$_2$ for x=0.28 and 0.36. The resulting ARPES intensities measured at several temperatures are shown for hole band in Fig. 4(a-d) and for electron band in Fig. 4(e-h); for x=0.36 the AF transition is fully suppressed and all data are taken in paramagnetic/tetragonal state. With increasing temperature, the hole band moves to higher binding energies, while the electron band moves to lower binding energies, similar to un-substituted BaFe$_2$As$_2$. As a result, the diameter of hole pocket decreases and electron pocket increases with increasing temperature. Therefore nesting present at low temperatures between hole and electron pocket weakens significantly upon warming up. To quantify this effect, we extracted the diameter of both pockets as a function of temperature by fitting the MDC's. Results are plotted in Fig. 5. At 50K the diameters of both the hole and electron pockets are close to 0.4~$\pi/a$, which corresponds to near nesting condition. Upon warming up to 300~K, the diameter of the hole pocket decreases to $\sim$0.32~$\pi/a$, while the diameter of the electron pocket increases to 0.48~$\pi/a$, which obviously destroys the nesting. We repeated the measurements, while cycling the sample temperature to ensure that this is an intrinsic effect and not due to aging of the sample surface. Similar data was also obtained for x=0.28 (not shown) demonstrating that this effect does not change with Ru substitution.  We should emphasize that, at these Ru substitution levels, the samples do not display AFM order at low temperatures despite presence of apparent nesting. The suppression of AFM order in these samples most is most likely due to magnetic dilution \cite{DhakaPRL11}. We note that the AFM nesting vector has a component along the z direction. We can compare size of both pockets at the same value of k$_z$ because the electron pocket is quasi 2D and its diameter does not change significantly along the z direction \cite{Chiang2D}.

\begin{figure}
\includegraphics[width=3.5in]{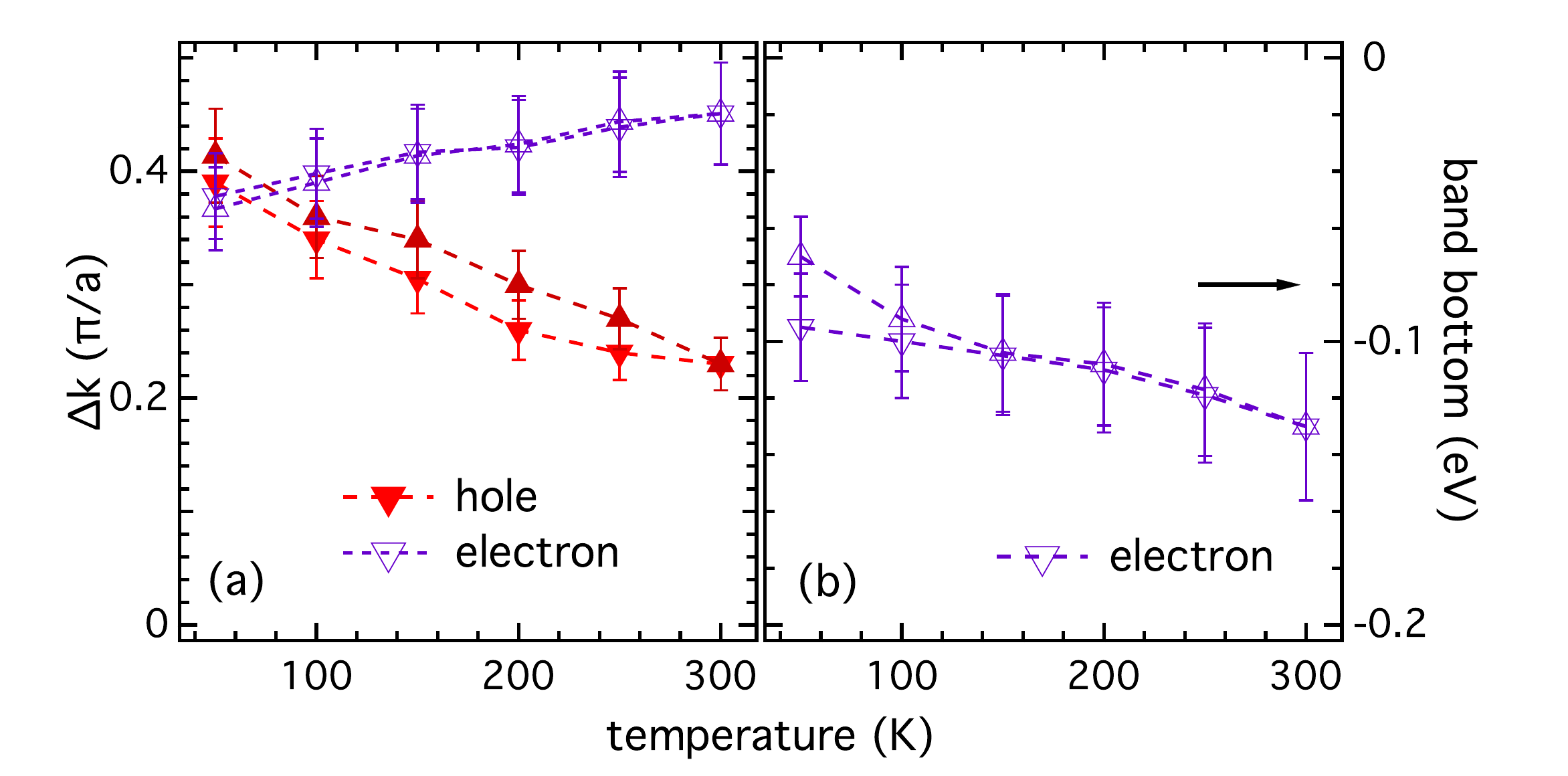}
\caption{(Color online) Dependence of the size of the hole and electron pockets on temperature for x=0.36 measured with h$\nu=40.8$~eV. 
(a) Diameter ($\Delta$k$_F$) of hole and electron pockets as a function of temperature, 
(b) energy position of the bottom of electron pocket as a function of temperature.}
\label{fig5}
\end{figure}

The large temperature induced changes of the band dispersion are indeed very unusual and unexpected. To the best of our knowledge, no such changes were reported in any of the studied systems before. The most obvious explanation of this effect is subtle changes in relative atomic positions due to thermal expansion. It is well known that the energy positions of the bands are very sensitive to the Fe-As distance. If the FeÐAs distance changes significantly with the temperature, this may explain our observations. We have therefore performed a band structure calculations using density functional theory (DFT) for the parent compound BaFe$_2$As$_2$ in order to understand the changes of the diameter of the pockets ($\Delta$k$_{\rm F}$) with temperature. We considered the tetragonal structure, and thermal expansion data\cite{Budko09}. Room temperature parameters were used for a and c. Next, we reduced a by the amount measured in the thermal expansion data. Initial calculations using the measured atomic positions show a band with d$_{x^2-y^2}$ character shifting downward by 17 and 15 meV for LDA and PBE exchange correlation functionals. This is approximately half the observed change (35$\pm$5 meV) from the ARPES measurements. A band with d$_{x^2-y^2}$ character shifts downward by about 5 meV. At the electron pocket, bands with d$_{xz}$/d$_{yz}$ character shift downwards by about 5 meV.

Using room temperature parameters for a and c, we allowed z$_{As}$ to relax such that all forces on the atoms equaled zero. We than reduced a by the amount measured in the thermal expansion data and again determined z$_{As}$ associated with this value of a. Near the Z ($\Gamma$) point the band with d$_{x^2-y^2}$ character shifts downward by 28 and 25 meV with LDA and PBE exchange-correlation functionals, respectively. This is fairly close to the observed change from the ARPES measurements. For the electron pocket the changes are small and almost no observable shift is present in the upper band (d$_{x^2-y^2}$ character), while, the lower bands have d$_{xz}$/d$_{yz}$ character and shift downwards by about 6 meV. This approach points to changes of the atomic positions due to thermal expansion as a possible explanation of our results. 

\begin{table}
\caption{\label{arsenic} Relaxed value of the internal arsenic position $z_{As}$ for different values of the lattice parameter $a$. $\Delta a/a$=0.0 and 0.6\% correspond to 0~K and 300~K, respectively.}
\begin{ruledtabular}
\begin{tabular}{llll}
$\Delta a/a$ & $z_{As}$ (LDA) & $z_{As}$ (PBE) & $z_{As}$ (exp.)\\ 
\hline
0.0\%&0.3413($\alpha=30.96^{\circ}$)&0.3452($\alpha=32.02^{\circ}$)&\\
0.6\%&0.3402($30.65^{\circ}$)&0.3442($31.75^{\circ}$)&0.3545($34.47^{\circ}$)\\
\end{tabular}
\end{ruledtabular}
\end{table}

\begin{figure}
\includegraphics[width=3.4in]{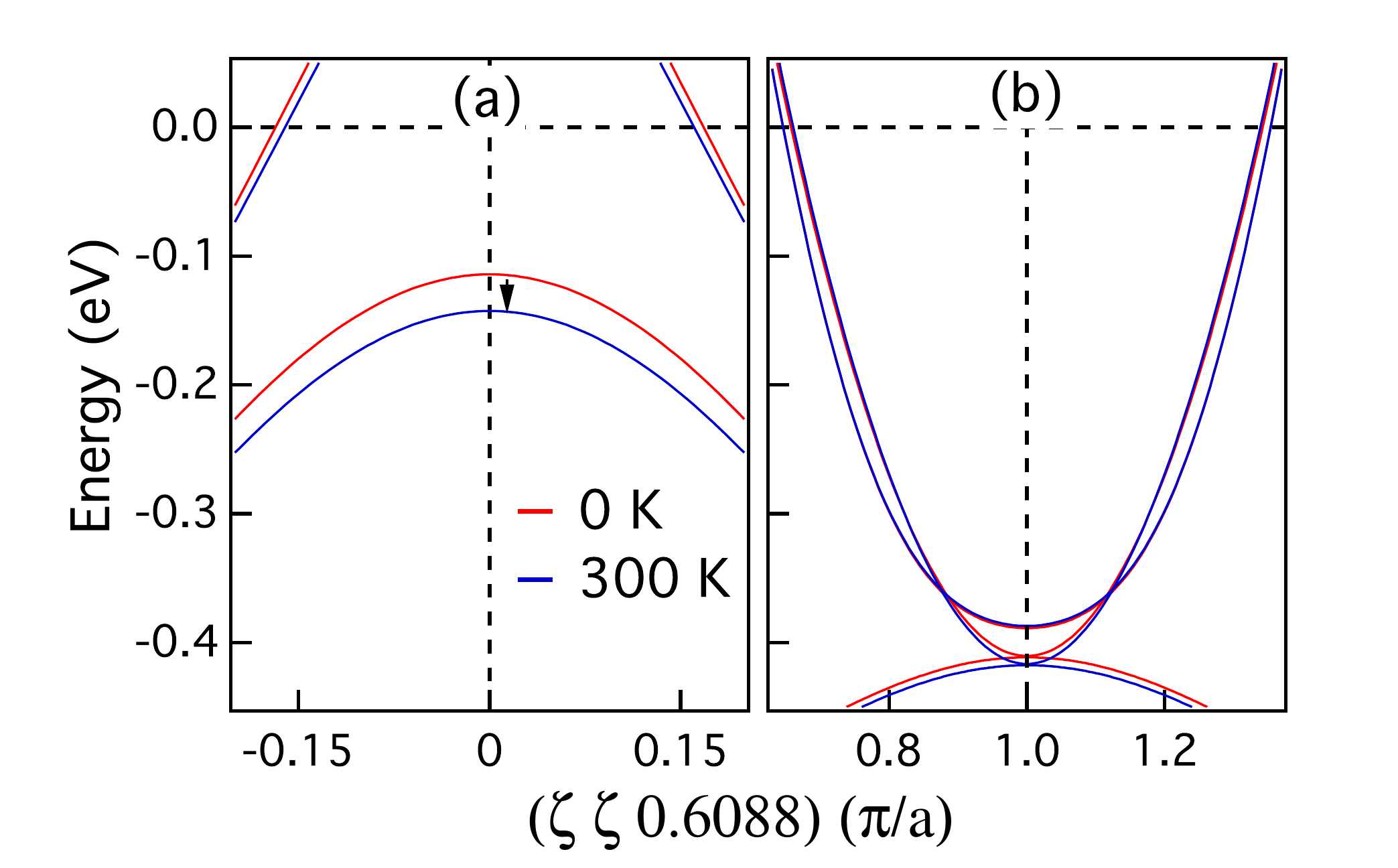}
\caption{\label{pockets} (Color online) Band structure calculation for BaFe$_{2}$As$_{2}$ (a) the $Z$  hole pocket and (b) the $X$ electron pocket.}
\end{figure}

In conclusion, we demonstrated that the band structure of pure and Ru substituted BaFe$_{2}$As$_{2}$ changes appreciably with temperature. The nesting conditions leading to SDW and present at low temperature become weaker upon warming up. This likely contributes to destruction of AFM order at higher temperatures. In essence, increasing temperature has similar effect on the band structure and nesting as Co substitution. This explains why the AFM/PM transition in the phase diagram is almost a straight line. At higher Co concentration, the nesting and AFM order is suppressed by smaller increase of the temperature. 

We would like to thank Rafael Fernandez, Andy Millis and Andrey Chubukov for very useful discussions, Sung-Kwan Mo at the ALS and staff at SLS for their excellent instrumental support. The work at the Ames Laboratory was supported by the Department of Energy, Basic Energy Sciences under Contract No. DE-AC02-07CH11358.

\end{document}